%% file: paper.tex
\documentstyle[12pt,rotating,epsfig,amsmath,here,amssymb,times,hhline]{article}
\newlength{\dinwidth}
\newlength{\dinmargin}
\begin{document}  
% The rest
\newcommand{\pom}{{I\!\!P}}
\newcommand{\reg}{{I\!\!R}}
\newcommand{\slowpi}{\pi_{\mathit{slow}}}
\newcommand{\pizero}{\pi^0}
\newcommand{\fiidiii}{F_2^{D(3)}}
\newcommand{\fiidiiiarg}{\fiidiii\,(\beta,\,Q^2,\,x)}
\newcommand{\n}{1.19\pm 0.06 (stat.) \pm0.07 (syst.)}
\newcommand{\nz}{1.30\pm 0.08 (stat.)^{+0.08}_{-0.14} (syst.)}
\newcommand{\fiidiiiful}{F_2^{D(4)}\,(\beta,\,Q^2,\,x,\,t)}
\newcommand{\fiipom}{\tilde F_2^D}
\newcommand{\ALPHA}{1.10\pm0.03 (stat.) \pm0.04 (syst.)}
\newcommand{\ALPHAZ}{1.15\pm0.04 (stat.)^{+0.04}_{-0.07} (syst.)}
\newcommand{\fiipomarg}{\fiipom\,(\beta,\,Q^2)}
\newcommand{\pomflux}{f_{\pom / p}}
\newcommand{\nxpom}{1.19\pm 0.06 (stat.) \pm0.07 (syst.)}
\newcommand {\gapprox}
   {\raisebox{-0.7ex}{$\stackrel {\textstyle>}{\sim}$}}
\newcommand {\lapprox}
   {\raisebox{-0.7ex}{$\stackrel {\textstyle<}{\sim}$}}
\def\gsim{\,\lower.25ex\hbox{$\scriptstyle\sim$}\kern-1.30ex%
\raise 0.55ex\hbox{$\scriptstyle >$}\,}
\def\lsim{\,\lower.25ex\hbox{$\scriptstyle\sim$}\kern-1.30ex%
\raise 0.55ex\hbox{$\scriptstyle <$}\,}
\newcommand{\pomfluxarg}{f_{\pom / p}\,(x_\pom)}
\newcommand{\dsf}{\mbox{$F_2^{D(3)}$}}
\newcommand{\dsfva}{\mbox{$F_2^{D(3)}(\beta,Q^2,x_{I\!\!P})$}}
\newcommand{\dsfvb}{\mbox{$F_2^{D(3)}(\beta,Q^2,x)$}}
\newcommand{\dsfpom}{$F_2^{I\!\!P}$}
\newcommand{\gap}{\stackrel{>}{\sim}}
\newcommand{\lap}{\stackrel{<}{\sim}}
\newcommand{\fem}{$F_2^{em}$}
\newcommand{\tsnmp}{$\tilde{\sigma}_{NC}(e^{\mp})$}
\newcommand{\tsnm}{$\tilde{\sigma}_{NC}(e^-)$}
\newcommand{\tsnp}{$\tilde{\sigma}_{NC}(e^+)$}
\newcommand{\st}{$\star$}
\newcommand{\sst}{$\star \star$}
\newcommand{\ssst}{$\star \star \star$}
\newcommand{\sssst}{$\star \star \star \star$}
\newcommand{\tw}{\theta_W}
\newcommand{\sw}{\sin{\theta_W}}
\newcommand{\cw}{\cos{\theta_W}}
\newcommand{\sww}{\sin^2{\theta_W}}
\newcommand{\cww}{\cos^2{\theta_W}}
\newcommand{\trm}{m_{\perp}}
\newcommand{\trp}{p_{\perp}}
\newcommand{\trmm}{m_{\perp}^2}
\newcommand{\trpp}{p_{\perp}^2}
\newcommand{\alp}{\alpha_s}

\newcommand{\alps}{\alpha_s}
\newcommand{\sqrts}{$\sqrt{s}$}
\newcommand{\LO}{$O(\alpha_s^0)$}
\newcommand{\Oa}{$O(\alpha_s)$}
\newcommand{\Oaa}{$O(\alpha_s^2)$}
\newcommand{\PT}{p_{\perp}}
\newcommand{\JPSI}{J/\psi}
\newcommand{\sh}{\hat{s}}
\newcommand{\uh}{\hat{u}}
\newcommand{\MP}{m_{J/\psi}}
\newcommand{\PO}{I\!\!P}
\newcommand{\xbj}{x}
\newcommand{\xpom}{x_{\PO}}
\newcommand{\ttbs}{\char'134}
\newcommand{\xpomlo}{3\times10^{-4}}  
\newcommand{\xpomup}{0.05}  
\newcommand{\dgr}{^\circ}
\newcommand{\pbarnt}{\,\mbox{{\rm pb$^{-1}$}}}
\newcommand{\gev}{\,\mbox{GeV}}
\newcommand{\WBoson}{\mbox{$W$}}
\newcommand{\fbarn}{\,\mbox{{\rm fb}}}
\newcommand{\fbarnt}{\,\mbox{{\rm fb$^{-1}$}}}
%
% Some useful tex commands
%
\newcommand{\qsq}{\ensuremath{Q^2} }
\newcommand{\gevsq}{\ensuremath{\mathrm{GeV}^2} }
\newcommand{\et}{\ensuremath{E_t^*} }
\newcommand{\rap}{\ensuremath{\eta^*} }
\newcommand{\gp}{\ensuremath{\gamma^*}p }
\newcommand{\dsiget}{\ensuremath{{\rm d}\sigma_{ep}/{\rm d}E_t^*} }
\newcommand{\dsigrap}{\ensuremath{{\rm d}\sigma_{ep}/{\rm d}\eta^*} }
% Journal macro
\def\Journal#1#2#3#4{{#1} {\bf #2} (#3) #4}
\def\NCA{\em Nuovo Cimento}
\def\NIM{\em Nucl. Instrum. Methods}
\def\NIMA{{\em Nucl. Instrum. Methods} {\bf A}}
\def\NPB{{\em Nucl. Phys.}   {\bf B}}
\def\PLB{{\em Phys. Lett.}   {\bf B}}
\def\PRL{\em Phys. Rev. Lett.}
\def\PRD{{\em Phys. Rev.}    {\bf D}}
\def\ZPC{{\em Z. Phys.}      {\bf C}}
\def\EJC{{\em Eur. Phys. J.} {\bf C}}
\def\CPC{\em Comp. Phys. Commun.}

\begin{titlepage}

\noindent
%Date:        26/05/2000               \\
%Version:     4.07 (Final)             \\
%Editors:     J. Stiewe,  M. Swart    \\
%Referees:    A. Bunyatyan, J. Olsson \\
%Comments to the editors and referees by Thursday, 27.04.2000
DESY 00-085  \hspace*{8.5cm} ISSN 0418-9833 \\
June 2000
\vspace{2cm}

\begin{center}
\begin{Large}

{\bf Inclusive Photoproduction of Neutral Pions \\
in the Photon Hemisphere at HERA}

\vspace{2cm}

H1 Collaboration

\end{Large}
\end{center}

\vspace{2cm}

\begin{abstract}
\noindent
The inclusive cross section for the
photoproduction of neutral pions has been measured as a
function of the transverse momentum, rapidity, and 
Feynman $x$ of the $\pizero$ mesons
at an average photon--proton centre-of-mass energy 
of $208$~GeV
and for photon virtualities below $Q^2=0.01$~GeV$^2$.
%The $\pizero$ mesons were reconstructed through their decays into two 
%photons which were detected in the electromagnetic 
%calorimeters of the H1 experiment.
The $\pizero$ measurement extends the range covered by previous charged 
particle measurements at HERA by two units of rapidity
in the photon direction 
down to a value of
$-5.5$ in the $\gamma p$ centre-of-mass frame.
% $-3.5$ in the laboratory frame.
The $\pizero$ transverse momentum distribution is well described over the whole
measured range by a power law ansatz, while an exponential fit falls
below the data at transverse momentum values above $1.5$~GeV/c.
Good agreement with the predictions of the Monte Carlo models PYTHIA and
PHOJET is found. 
In the context of the PYTHIA model the data are inconsistent with
large intrinsic transverse momentum values in the photon.
\end{abstract}
\vspace{1.5cm}

\begin{center}
To be submitted to European Physics Journal  
\end{center}

\end{titlepage}

%
%          COPY THE AUTHOR AND INSTITUTE LISTS 
%       AT THE TIME OF THE T0-TALK INTO YOUR AREA
%
% from /h1/iww/ipublications/h1auts.tex 
% from /h1/iww/ipublications/h1inst.tex

\include{h1autsn} % contains h1auts.tex and h1inst.tex

\newpage
\section{Introduction}

\noindent
At the electron-proton collider HERA, photoproduction ($\gamma p$)
processes are induced by quasi-real photons
that are emitted by the incoming electron or 
positron\footnote{In the running periods used for this analysis, 
HERA was operated with a positron beam.}.
The majority of the photon-proton interactions are ``soft'' scattering
processes in which particles with limited transverse momenta are produced.
A small fraction of $\gamma p$ processes, however, involve
``hard'' scatterings. In these, the photon interacts either as a 
single object with a parton from the proton (``direct'' photon
interactions) or via a parton-parton scattering after fluctuating
into a partonic system.
These latter ``resolved'' photon processes give rise to a photon remnant
following the direction of the incoming photon.

In the study of photoproduction scattering dynamics 
inclusive spectra of identified particles constitute an important source
of information.
Investigations of inclusive spectra using charged particles at 
HERA have confirmed the general features of hard photon-proton 
scattering outlined above. In particular it was 
shown that in these photoproduction processes 
there is an excess of charged particles in the region of large
transverse momenta, as compared to hadron-hadron scattering processes,
and that this excess is in 
agreement with perturbative Quantum Chromodynamics (QCD) 
calculations, which take into account both the direct
processes and the hard components of the resolved photon processes.  
These studies
were carried out using the central tracking detectors of the 
experiments H1~\cite{h1cha,h1cha2} and ZEUS~\cite{ZEUS}.  
The photon hemisphere was explored for charged particles
down  to $-1.5$ units of pseudorapidity $\eta$ in the laboratory
      system\footnote{The pseudorapidity and rapidity are defined as
      $\eta = -\ln \tan(\theta/2)$ and 
      $y = \frac{1}{2}\ln\frac{E+p_z}{E-p_z}$, respectively, 
      where $E$ is the particle energy, $p_z$ the momentum
      component with respect to the $z$-axis and $\theta$ the polar angle;
      the $z$-axis is given by the proton direction.} 
and with transverse momenta up to $12$ GeV/c. 

           The $\pi^0$ measurement described here
           is an extension of the previous charged particle studies 
           into  an as yet unexplored phase
           space region in which 
           the photon remnant is expected to dominate
           the particle flow. 
           In this region  charged particles
           could not be reliably reconstructed, because of 
           limitations in acceptance of the H1 tracking devices.
%           can no longer be
%           reliably recorded by the H1 detector, owing to limitations
%           in acceptance of the tracking devices. 
           Thus, the present $\pi^0$ measurement covers     
           two additional units of rapidity 
           $y$, down to $y=-3.5$. We
           note that, since at HERA the $\gamma p$  centre-of-mass (CM)
           frame itself 
           moves with an average rapidity 
           of roughly two units relative to the laboratory 
           system, this limit corresponds to 
           $y_{\gamma p}\simeq -5.5$, as seen in the $\gamma p$ CM frame.

The production
of neutral pions has previously been extensively studied
in several fixed target experiments \cite{fixedtarget, Omega}, 
using both charged particle beams and photon beams,
as well as in $pp$ and $p\bar{p}$ 
collider experiments \cite{ISR}.
The CM energies in these photoproduction
experiments range up to 18~GeV. In this paper we study, for the first
time at HERA, neutral pion photoproduction at
$\gamma p$ CM energies around $208$~GeV. 
We measure the inclusive double differential
$\pizero$ photoproduction cross section as a function of the $\pizero$
transverse momentum and rapidity in the laboratory frame.
The $\pizero$ mesons are identified through their decay into two 
photons which are detected in the electromagnetic calorimeters of the
H1 detector.
%The large acceptance of these calorimeters together with their fine 
%granularity and high resolution gives access to the laboratory rapidity 
%values quoted above, and to transverse momenta up to $2$~GeV/c.

The paper is organized as follows.
In section~2, we give a short description of the H1 detector components
that are most relevant to the $\pizero$ cross section measurement.
In section~3, we describe the Monte Carlo models used to correct the data.
Section~4 outlines the data selection and the analysis method, while the
systematic errors are discussed in section~5. 
The results are presented and discussed in section~6. Finally, a summary
is given in section~7.

\section{Detector Description}

A comprehensive description of the H1 detector is given 
in Ref.~\cite{h1det}. 
The parts of the detector vital for this analysis are
the backward calorimeter (SpaCal),
the Liquid Argon calorimeter (LAr), the central trackers 
and the electron and photon detectors of the luminosity system.
Their main properties are described here.

The SpaCal \cite{spacal} is a lead-scintillating fibre calorimeter 
separated into an electromagnetic (EM) 
and a hadronic section of equal
size. The EM section consists of 1192 cells with a cross section of 
$40.5 \times 40.5 \;\text{mm}^2$, each read out by a photomultiplier tube.
The hadronic section consists of 128 cells with a cross section
of $120 \times 120 \;\text{mm}^2$.
The EM section has a depth of 27 radiation lengths.
The polar angle range
covered by the SpaCal is
$153^\circ \le \theta \le 178^\circ$.

The energy resolution of the EM SpaCal section for electromagnetically
interacting particles is 
\begin{equation} 
\frac{\sigma(E)}{E} \approx 
   \frac{0.075}{\sqrt{E/\text{GeV}}} \oplus 0.010  ,
\end{equation}
where E is the energy deposited in the SpaCal. 
The absolute energy scale,
in the lower energy range from 0.2~GeV to 10~GeV,
is known with an uncertainty of $4\%$ \cite{Martin}.
The resolution in $\theta$ 
is better than 2.5 mrad for energies above 1~GeV.

The LAr calorimeter \cite{calo} is highly segmented and
consists of an electromagnetic section with lead absorbers, corresponding
to a depth varying between 20 and 30 radiation lengths, and a hadronic section
with steel absorbers. The resolution of the LAr EM section is 
\begin{equation} 
\frac{\sigma(E)}{E} \approx \frac{0.12}{\sqrt{E/\text{GeV}}} \oplus 0.01 ,
\end{equation} 
as measured in test beams \cite{larcal}.
Furthermore the absolute energy scale of the LAr calorimeter,
in the low energy range from 0.8~GeV to
10~GeV, is known to a precision of $4\%$.
The LAr calorimeter covers the polar angle
range $4^\circ \le \theta \le 154^\circ$.
 
The H1 central tracking system \cite{h1det} is mounted concentrically
around the beam line and
covers polar angles in the range $20^\circ < \theta < 160^\circ$. 
Measurements of the momenta of charged particles are provided by
two coaxial cylindrical drift chambers (central jet chambers, CJC), mounted
inside a homogeneous magnetic field of 1.15 Tesla.
Further drift chambers are placed radially inside and outside the
inner CJC chamber, providing
accurate measurements of the $z$ coordinates of charged tracks. Finally,
multiwire proportional chambers (MWPC), which allow triggering on those
tracks, are also located within and between the CJC chambers.
In the present analysis tracking information is used to identify
isolated electromagnetic clusters in the LAr calorimeter, and to find
the interaction vertices of the selected events.
 
To detect photoproduction events, the 
electron and photon detectors of the H1 luminosity 
system \cite{h1det} are used. They
consist of TlCl/TlBr crystal calorimeters with an energy resolution of
$ \sigma(E)/E = 0.22/\sqrt{E/\text{GeV}}$ and are located at $z = -33$~m
and $z = -103$~m, respectively. The detectors serve to identify
scattered positrons from photoproduction processes.
The signature of photoproduction is a signal in the electron
detector with no accompanying signal in the photon detector.

\section{Event Simulation}

In this analysis, photoproduction interactions are modelled  
by two Monte Carlo generators, PHOJET (version 1.04)
\cite{phojet} and PYTHIA (version 5.722) \cite{pythia}. 
These have both been shown to describe photoproduction
data in previous analyses \cite{h1cha, h1cha2, Rick}.
Both generators have in common that they use leading order (LO) QCD
matrix elements for the hard scattering subprocess. Initial and final
state parton radiation and the LUND fragmentation model \cite{lund}
for hadronization
are included, as implemented in the JETSET program \cite{jetset}.
The programs differ, however, in the treatment of multiple 
interactions and the transition from hard to soft processes at low
transverse parton momentum.

The PHOJET event generator simulates all components
that contribute to the total photoproduction cross section. It is
based on the two-component dual parton model (for a review, see \cite{DPM}).
PHOJET incorporates simulations of soft processes on the basis of Regge
theory. Soft and hard processes are connected through a unitarization
scheme which serves to keep cross sections finite.

The PYTHIA event generator uses LO QCD calculations for the primary
parton-parton scattering process and for multiple parton interactions.
The latter are considered to arise from the scattering of partons from 
the photon and proton remnants.
PYTHIA also models both hard and soft hadronic interactions - the latter
by the exchange of low energy gluons - applying
a unitarization scheme \cite{Schuler}.

   Additional transverse momentum is generated in PYTHIA
   by virtue of the assumed intrinsic (primordial) transverse momentum    
   ($k_\perp$) of the partons in the
        interacting hadrons. The $k_\perp$ distribution of the quark 
        and antiquark from the photon, fluctuating into a hadronic state,
        is parametrized as a power law function
        $1/(k_{\perp 0}^2 + k_\perp^2)$ \cite{Schuler} with the value
        $k_{\perp 0} = 400$~MeV/c. 
For this function PYTHIA applies for the primordial $k_\perp$
a cutoff,
the value of which is taken from the transverse momentum
of the individual hard scattering subprocess.
        PYTHIA offers alternative parametrizations for the
        $k_\perp$ distribution, by either a Gaussian or an exponential 
        function. The predictions of the model
        for different
%%%        choices of the 
        $k_\perp$ distributions
%%%        and for
%%%        different values of the parameters defining 
%%%        the shape and strength of the $k_\perp$ distribution,
        are compared with the experimental results in section~6.

%%  PHOJET comment about primordial k_t can be added here
The PHOJET 1.04 event generator has no explicit intrinsic 
transverse momentum for partons entering a hard scattering.

For both Monte Carlo models the factorization and renormalization scales
were set to the transverse momentum of the final state partons. GRV-LO
\cite{grv} parton distribution functions for the proton and 
photon were used for the generation of events.

\section{Data Selection and Analysis}

Photoproduction events were selected, which have
the scattered positron registered in the electron detector 
of the luminosity system and
consequently have a four-momentum transfer $Q^2 < 10^{-2}~{\rm GeV}^2$,
where $Q^2 = 4 E_eE^{\prime}_e \cos^2(\theta/2)$. Here $E_e$ and 
$E_e^{\prime}$ are the incoming and scattered positron energies, and $\theta$
the polar angle of the scattered positron, $\theta \sim 180^\circ$.
More specifically, the events
were required to have a positron candidate registered 
in the fiducial volume of the
electron detector, with energy $E^{\prime}_e >4$~GeV, and to have less than
$2$~GeV deposited in the photon detector. The latter condition suppresses
background from the proton beam appearing in coincidence with the high rate
of events from the Bethe-Heitler reaction $e p \rightarrow e \gamma p$.
%; it also reduces the QED corrections to the cross sections.
In order to ensure full efficiency and acceptance of the photoproduction
tagging condition, the range of the variable 
$y_{\rm B}=1-(E^{\prime}_e/E_e)$ is restricted to $0.35<y_{\rm B}<0.65$
in the present analysis. 

Two data samples were used in the analysis:
\begin{itemize}
\item
The ``minimum bias'' data sample, collected in 1997 and corresponding to 
an integrated luminosity of $\sim300~{\rm nb}^{-1}$. In addition to the 
photoproduction tagging condition, the 
trigger demanded only loose conditions on the 
presence of charged tracks in the central tracking detectors.
\item
The ``SpaCal'' sample, collected in 1996 and corresponding to
an integrated luminosity of 
 $\sim4.3~{\rm pb}^{-1}$. The trigger  used in this sample contained,
in addition to the photoproduction tagging condition,
 requirements on the energy
registered in the EM SpaCal.
Thus at least one SpaCal energy cluster 
(i.e. an isolated energy deposition, separated
from neighbouring energy depositions by the clustering algorithm)
had to exceed $2$~GeV, and at the same time have a radial distance to the
beam axis of more than 16~cm.
In the subsequent analysis a trigger efficiency of at least $40~\%$ 
was ensured by demanding that in each event  
the most energetic SpaCal cluster had an energy exceeding 
$2.2$~GeV and a 
radial distance to the beam axis exceeding 16~cm.
Full trigger efficiency was reached for cluster energies
and radial distances larger than $2.6$~GeV and $22$~cm, respectively.
\end{itemize}

The ``SpaCal'' sample allows the study of the phase space 
region in which the number of events in 
the ``minimum bias'' sample is too small, i.e. the region where 
the $\pizero$ has both large negative rapidity and large transverse momentum.

In the further event selection the 
$z$-coordinate $z_v$ of the event interaction vertex,
reconstructed using the central tracking detectors,
had to satisfy the condition $|z_v|<35$~cm.
The cut on $z_v$ suppresses beam-gas background events.

The resulting ``minimum bias'' and ``SpaCal'' data samples 
consist of $\sim~115$K events and $\sim~500$K events, respectively.

In the subsequent $\pizero$ reconstruction, additional conditions were applied
in order to accept energy clusters in the LAr and SpaCal EM calorimeters as 
photon candidates.
\begin{itemize}
\item
For a LAr cluster to be accepted as a single photon candidate, 
the transverse radius of the cluster had to be smaller than 8 cm
and the longitudinal cluster extension below 10~cm. 
The cluster energy had to exceed $0.3$~GeV.
%A lower limit on
%the single cluster energy  was implicitly
%given  by the lower $p_\perp$ bound of $0.8$~GeV/c required for the
%reconstructed $\gamma\gamma$ combinations. Here, 
%$p_\perp$ is the transverse momentum with respect to
%the beam line, calculated from the sum of the
%momentum vectors of the single photons.
Furthermore,
the distance between the cluster position and the impact point of the
closest reconstructed charged track had to be larger than 5~cm.
\item 
An EM SpaCal cluster was associated with a photon if it was
reconstructed as an isolated energy deposition covering at least two
SpaCal cells.
The cluster energy had to exceed 0.3~GeV. A radial distance between 8~cm and
75~cm from the cluster to the beam axis was required to ensure
full energy containment. Information from the backward tracking detectors
was not used in selecting photon candidates in the SpaCal, since many photons
shower in the material in front of the backward tracking detectors and
the SpaCal. 
\end{itemize}

\begin{figure}[H]
  \vspace*{-0.5cm}
  \begin{center}
    \begin{tabular}{cc}
      \epsfig{file=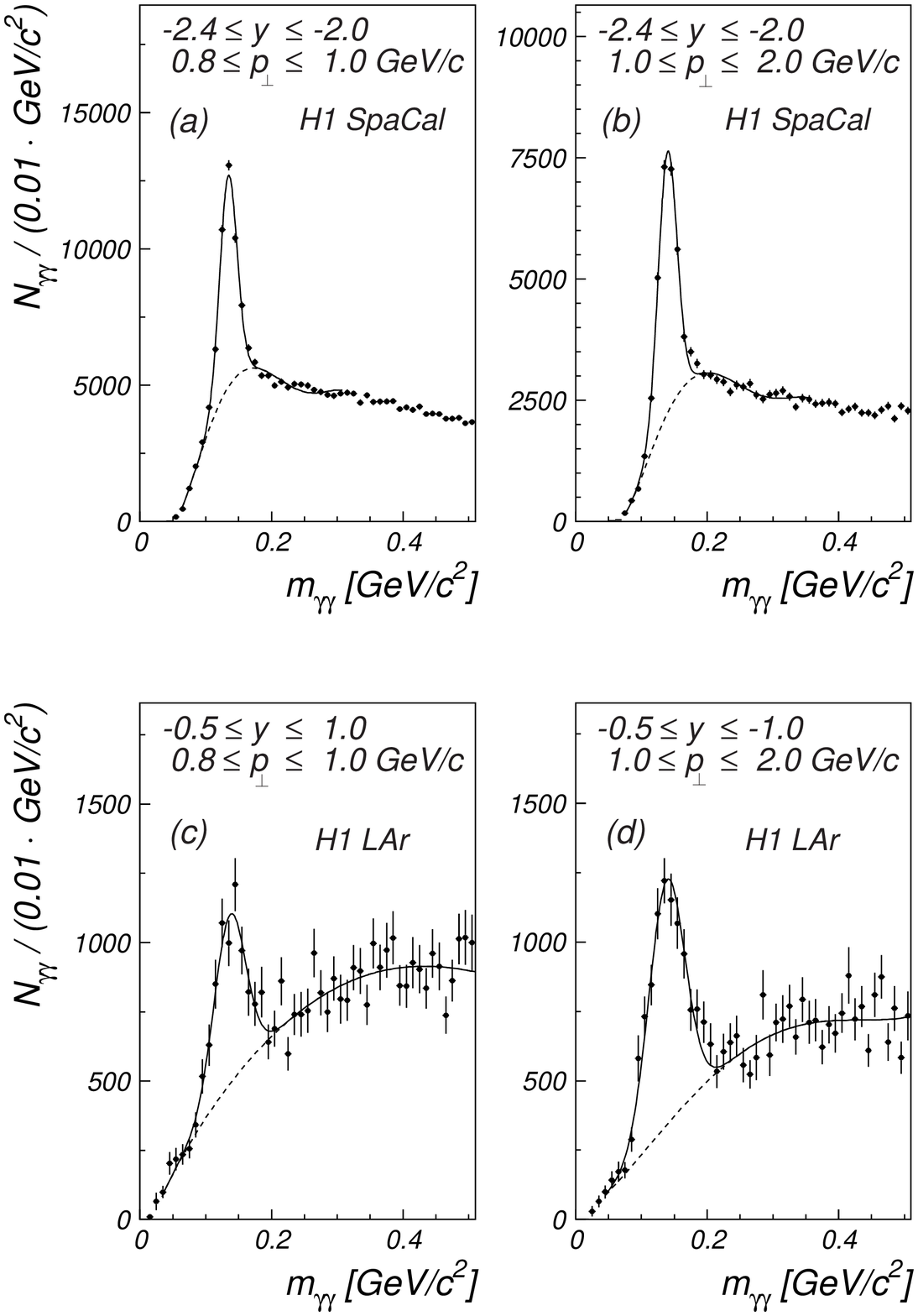,width=0.7\textwidth}
    \end{tabular}
    \parbox{16cm}{\caption{\label{lego} Two--photon invariant mass
        distributions reconstructed in the
        SpaCal calorimeter (a, b) and in the Liquid Argon calorimeter (c, d).
        The full curves show fits of a sum of a Gaussian and a background 
         distribution (dashed curves) as described in the text. }}
  \end{center}
\end{figure}

The procedure for reconstructing and counting $\pizero$ mesons was as follows: 
Each photon pair, from either the LAr or the SpaCal and
satisfying the above criteria, was labelled according to transverse momentum 
$p_\perp$ and rapidity $y$.
Photon four-momenta were derived from the cluster
energy, the cluster coordinates and the coordinates of the interaction vertex.
For every photon pair, the invariant mass $m_{\gamma \gamma}$ was calculated
according to
\begin{equation} 
m_{\gamma \gamma}^2 c^4 = 2 E_1 E_2 (1 - \cos \psi_{12}) \; .
\label{mass}
\end{equation}
\noindent
Here $E_1$ and $E_2$ are the energies of the measured calorimeter
clusters, attributed to the decay photons, and $\psi_{12}$ is
their opening angle.
Fig.~\ref{lego} shows two--photon invariant mass distributions in
several regions of transverse momentum 
$p_\perp$ and rapidity $y$ of photon pairs measured in the SpaCal
(Fig.~\ref{lego}a and~\ref{lego}b), and in the LAr calorimeter
(Fig.~\ref{lego}c and~\ref{lego}d). These variables were
calculated from the four-momentum sum of the two photons. 

In each bin of $p_\perp$ and $y$ the two-photon mass distribution can be
satisfactorily described by the sum of a Gaussian function for the $\pi^0$ 
signal and a background curve.
The background was determined by fitting a polynomial of
fourth order multiplied by a function of the type 
($m_{\gamma\gamma} - m_{\gamma \gamma}^{thr})^{\alpha}$ in order
to improve the threshold description where the background rises
steeply. $m_{\gamma \gamma}^{thr}$ is 
given by the centre of the lowest non-empty
interval in the mass distribution, 
and $\alpha$ is a fitted parameter with $0 \le \alpha \le 1$.

The double differential cross section for inclusive
$\pi^0$ photoproduction is given by the expression 
\begin{equation}
\frac{d^2 \sigma_{\gamma p}}{d p_\perp^2 d y} =
\frac{N_{\rm prod} (\Delta p_\perp,\Delta y)}
 {2p_\perp \cdot \Delta p_\perp 
\cdot \Delta y \cdot \Phi \cdot \cal L}.
\end{equation}

\noindent
$N_{\rm prod}$ is the number of $\pi^0$s produced in the transverse
momentum ($\Delta p_\perp$) and rapidity ($\Delta y$) interval considered, 
determined with all efficiency and acceptance corrections. 
The efficiency was determined by simulation of photoproduction events
using the PHOJET and PYTHIA generators (see section 3), with full
simulation of the detector response.
$\cal L$ is the integrated luminosity, and $\Phi =$ 0.00968 is the photon flux
factor, calculated according to the Weizs\"acker - Williams formula
\cite{WeWi}
for the $y_{\rm B}$ range covered in this analysis.
This range 
corresponds to the $\gamma p$ CM energy range 
$177<\sqrt{s_{\gamma p}}<242$~GeV, 
with an average of $208$~GeV. We note
that in the various bins of $p_\perp$ and $y$ the 
 average $\sqrt{s_{\gamma p}}$ may deviate from this value
 by up to $\pm 3\%$.

%Control distributions of kinematical quantities concerning the 
%relevant triggering
%and event 
%selection criteria for data and simulation show very good agreement.

%The procedure was cross checked with the help of the PYTHIA generator. 

\section{Systematic Errors}

The errors on the results are dominated by
systematic effects with the following main sources:

\noindent
1) The uncertainty of the fitting procedure arising from the determination
of the combinatorial background in the fits of the two-photon mass
distributions.
%1) The systematic error on the numbers of $\pi^0$s due to the uncertainty
%in the 
%determination of the combinatorial background in the  
%two-photon mass distributions. 
In all kinematical intervals relevant for
this analysis this was estimated by fitting these distributions,
varying the fit 
parameters and fit ranges several times, following a random
choice. Also the order of the polynomial used to describe the 
background was varied. 
%For each interval the maximum and minimum
%number of $\pi^0$s, $N_{\pi^0}^{max}$ and $N_{\pi^0}^{min}$, were thus 
%obtained.
%The number of $\pi^0$s produced in each interval, together with 
%the associated uncertainty, was then determined as
%$N_{\pi^0} = $1/2$ \; \cdot \; (N_{\pi^0}^{max} + N_{\pi^0}^{min})$ and
%$\Delta(N_{\pi^0}) = N_{\pi^0} - N_{\pi^0}^{min}$, respectively.
The errors derived in this way vary substantially due to the
%accidental behaviour of the background.
statistical uncertainty of the background estimation.
In order to get a conservative but reliable estimate for the uncertainty 
from the fitting procedure, a global error of $8$\% was therefore assigned,
based on the distribution of the errors.

\noindent
2) The $\pizero$ reconstruction efficiencies, derived using the PYTHIA
and PHOJET models, are in general in good agreement, except in the 
bins of high $p_\perp$ where the two models differ by up to $30$\%. 
Such differences can be understood as being due to internal differences 
of the models in the population of certain kinematical regions. 
In the cross section calculations we have 
taken the mean reconstruction efficiency and assigned the difference
as a bin to bin systematic error, varying between $1$\% and $15$\%.

%2) The model dependence of the $\pi^0$ reconstruction
%efficiency, determined by comparing
%the results derived from PYTHIA and PHOJET, leads to systematic
%uncertainties varying from bin to bin between $1~\%$ and $15~\%$.

\noindent
3) The uncertainty of the energy scales for the LAr and
SpaCal calorimeters: Shifting the scale upward (downward) by $4$\%
increases (reduces) the cross sections by about $20$\% for the data
selected with the SpaCal trigger, and by about $10$\% for the minimum
bias triggered data.

\noindent
4) The uncertainty on the electron detector acceptance is $4$\%
and $6$\%, for the 1996 and 1997 data samples, respectively.

\noindent
5) The uncertainty on the efficiency of the SpaCal trigger  
is found to be $1$\%, from cross checks with an independent trigger.

\noindent
6) The contribution of $\pi^0$s from beam-gas interactions
has been estimated to be less than $1$\%
using the so-called ``pilot bunches'', i.e. positron beam bunches 
without partners with which they can collide in the proton beam.

\noindent
7) The integrated luminosity is known to a precision of 
better than $2$\%.

\section{Results}

\begin{figure}[H]
  \begin{center}
    \begin{tabular}{cc}
      \hspace*{-0.5cm}
      \epsfig{file=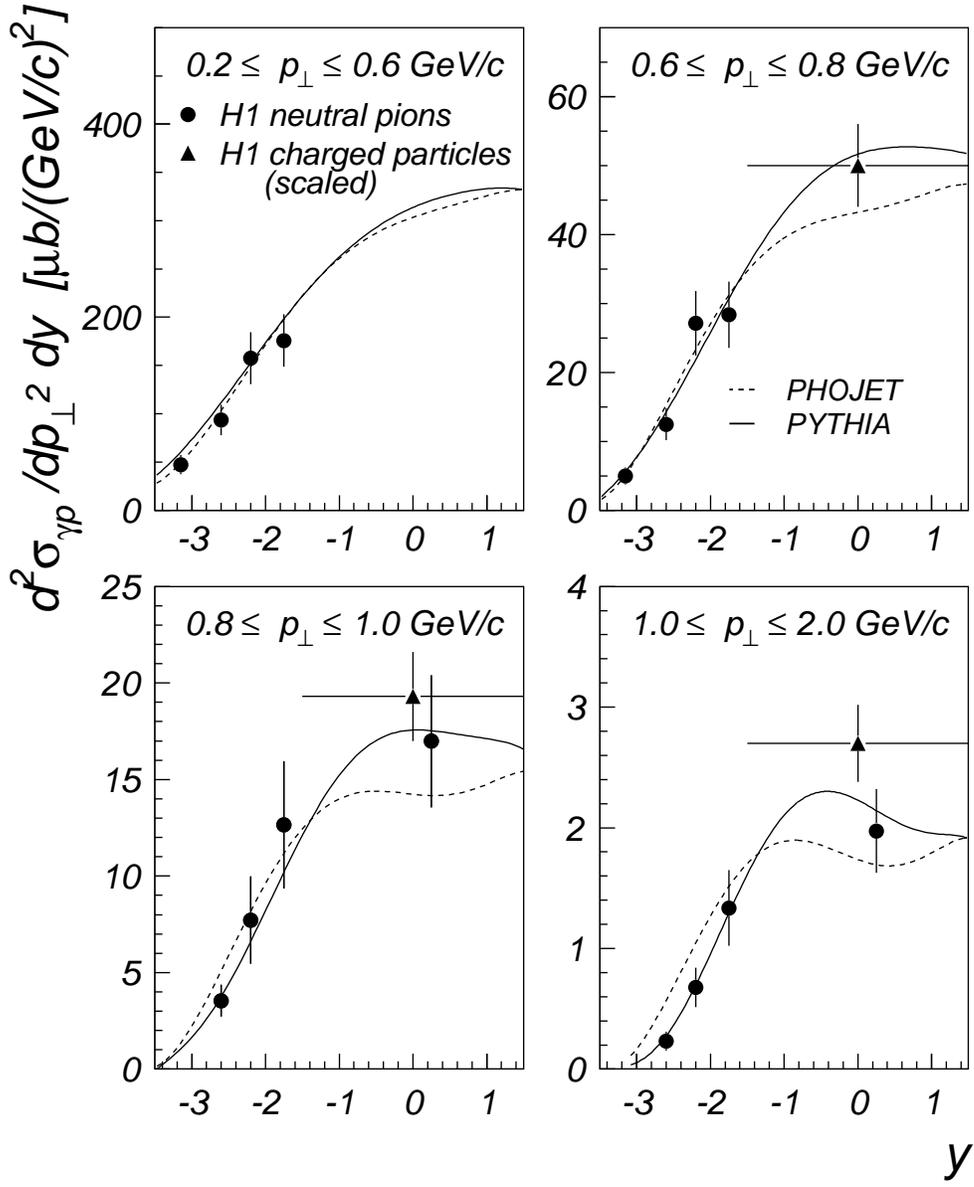,width=0.8\textwidth}
    \end{tabular}
    \parbox{16cm}{\caption{\label{dsy} 
        Inclusive $\pizero$ photoproduction cross 
        section as a function of laboratory rapidity 
        $y$ in intervals of transverse momentum $p_\perp$ (full circles).
        The triangles are the corresponding cross section
        values for charged pions, derived from \cite{h1cha} by
        subtracting a fraction of $17.5$\% to account for the
        ``non-pion'' contribution, and then dividing by 
        the isospin factor 2 (see text).
        The curves are the predictions of the PHOJET (dashed) 
        and PYTHIA (full) event generators for neutral
        pions. 
}}
  \end{center}  
\end{figure}

\begin{figure}[H]
  \begin{center}
    \begin{tabular}{cc}
      \hspace*{-0.0cm}
      \epsfig{file=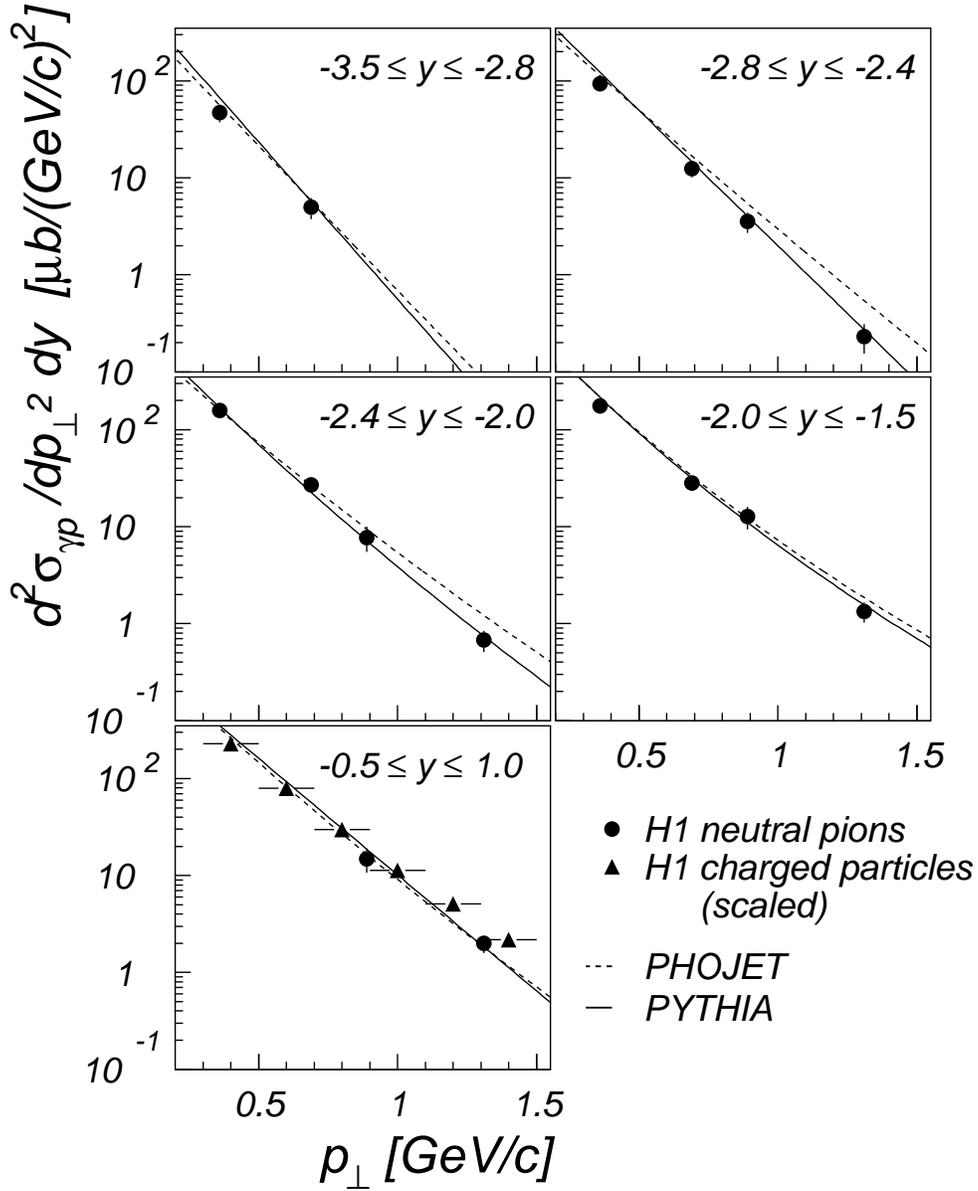,width=0.8\textwidth}
    \end{tabular}
    \parbox{16cm}{\caption{\label{dspt}   
        Inclusive $\pizero$ photoproduction cross 
        section as a function of transverse momentum $p_\perp$
        in intervals of laboratory rapidity $y$ (full circles).
        The triangles are the corresponding cross section
        values for charged pions in the pseudorapidity range
        $|\eta| \le 1.5$, derived from \cite{h1cha} by
        subtracting a fraction of $17.5$\% to account for the
        ``non-pion'' contribution, and then dividing by 
        the isospin factor 2 (see text).
        The curves are the predictions of the PHOJET (dashed) 
        and PYTHIA (full) event generators for neutral pions.
}} 
\end{center}
\end{figure}

The cross section for inclusive neutral pion production in 
$\gamma p$ interactions is displayed in Fig.~\ref{dsy} 
as a function of the $\pi^0$ rapidity for four intervals of
the transverse momentum and in Fig.~\ref{dspt} as a function
of $p_\perp$ for five intervals of rapidity.
The data are summarized in Table~1. 
Rapidity is measured in the laboratory frame.
Bin-centre corrections have been applied and
errors are quoted as the sum of statistical and systematic errors, 
added in quadrature. In regions of low statistics 
(low $y$, high $p_\perp$) or large combinatorial background
(central region, low $p_\perp$) unstable fits prevent 
the determination of cross section values.

Values derived from the H1 charged particle cross section 
measurements~\cite{h1cha}, which 
cover the pseudorapidity range $|\eta| \le 1.5$,
are also shown (full triangles) in Figs.~\ref{dsy} and~\ref{dspt}.
In order to compare these measurements
with the $\pi^0$ result, 
the charged 
particle cross sections were first corrected 
for the non-pion part
%to $\pi^+ + \pi^-$ cross sections
by subtracting a fraction of $(17.5 \pm 0.2)$\% as derived from the
PYTHIA and PHOJET predictions (the error reflects the difference in 
the predictions). The resulting $\pi^+ + \pi^-$ cross sections
were then further divided by the isospin factor $2$.
The $e p$ cross sections
given in \cite{h1cha} have been converted into $\gamma p$ 
cross sections using the appropriate flux factor. 
While the charged pion cross section entries in Fig.~\ref{dspt}
correspond directly to the original data points in \cite{h1cha}, 
a smooth interpolation was used to arrive at the charged
pion cross sections in Fig.~\ref{dsy}, 
in order to account for the different transverse
momentum ranges. 
%To account for the different transverse momentum ranges,
%the charged particle cross sections in Fig.~\ref{dsy} 
%were derived with the help of a smooth  
%interpolation of the charged particle differential
%cross section  w.r.t. transverse momentum, as published in \cite{h1cha
%The charged cross section entries in Fig.~\ref{dspt} represent the 
%original data points from the paper quoted. 
The agreement between the neutral and charged pion cross sections is good 
in the kinematical range covered by both analyses. Figs.~\ref{dsy} 
and~\ref{dspt} 
demonstrate well the additional phase space region accessed with 
the present neutral pion analysis. 

The model predictions according to the PHOJET and PYTHIA simulations 
are also shown in Figs.~\ref{dsy} and~\ref{dspt}, with satisfactory 
agreement especially for PYTHIA. PHOJET predicts slightly too 
large a $\pizero$ rate at high $\trp$ and low $y$.

\begin{figure}[H]
  \begin{center}
    \begin{tabular}{cc}
      \hspace*{-0.5cm}
     \epsfig{file=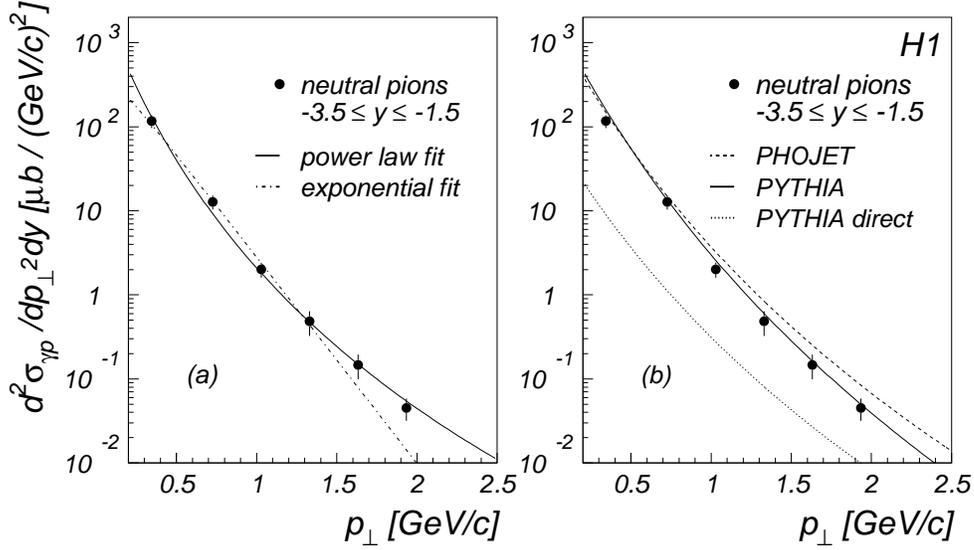,angle=270,width=0.8\textwidth}
%\begin{turn}{270}
%      \epsfig{file=pi0_fig4.eps,width=0.8\textwidth}
%\end{turn}
    \end{tabular}
    \parbox{16cm}{\caption{\label{dsint} 
        (a) Inclusive $\pizero$ photoproduction cross section as a 
        function of transverse momentum $p_{\perp}$ (circles), for 
        the rapidity region $-3.5\le y\le -1.5$.
%  for $Q^2 \le 0.01$~GeV$\,^2$ and $0.35 \le y_{Bj}\le 0.65$.
        The dash - dotted curve is an exponential fit
        to the data in the range $0.3 \le p_\perp \le 1.6$~GeV/c;
        the full curve is the result of a power law fit
        to all data points (see text). (b) Same cross section as in (a) 
        compared to predictions of PHOJET and PYTHIA. In addition,
        the contribution of direct photon interactions in PYTHIA is shown.
        }} 
  \end{center}
\end{figure}

The  $\pizero$ transverse momentum spectrum, integrated over the
laboratory rapidity range covered by the SpaCal, i.e. 
$ -3.5 \le y \le -1.5 $, is shown in  Fig.~\ref{dsint}. The data
are summarized in Table $2$. In deriving these cross sections also
those regions were included for which no results are given in Table 1.

Transverse momentum spectra in high energy hadron - hadron collisions
are successfully described by a power law ansatz of the form
\begin{equation} 
\frac{d^2 \sigma}{d p_\perp^2 dy} = 
           A \cdot (1 + p_\perp/p_{\perp 0})^{-n} \; .
\end{equation}

\noindent
This QCD inspired ansatz \cite{Blanken,hagedorn} was designed to describe
transverse momentum spectra of centrally produced particles.
It was shown in \cite{h1cha,h1cha2,ZEUS} that it 
fits well the H1 and ZEUS spectra measured for charged
particles in the central pseudorapidity range. 
This is in agreement with
observations made in many hadron-hadron experiments, 
see e.g. Kourkoumelis et al.~\cite{ISR} 
for centrally produced $\pi^0$s at various
centre-of-mass energies at the CERN ISR, and Banner et al.~\cite{ISR} for
measurements performed at $\sqrt{s} = 540$~GeV at the CERN SPS. A review
of large $\trp$ particle production at the ISR is given by 
Geist et al.~\cite{Geist}.

\vspace*{0.5cm}
{\small \begin{table}[h]
  \begin{center}
    \begin{tabular}[h]{|l||c|c|c|c|} \hline
      & \multicolumn{4}{|c|}{ $p_{\perp}$ [GeV/c] } \\ \cline{2-5}
      \multicolumn{1}{|c||}{{ \raisebox{0.0ex}[-2.5ex]{$y$} } }
      &{ $[0.2,0.6]$ } 
      &{ $[0.6,0.8]$ }
      &{ $[0.8,1.0]$ }
      &{ $[1.0,2.0]$ }  
      \\ 
      & (0.36)
      & (0.69)
      & (0.89)
      & (1.31)
      \\ \hline \hline
      { $[-3.5,-2.8] \, (-3.15)$} & $47\pm10$   & $5.0\pm1.2$  &   -      &    -
      \\ \hline 
      { $[-2.8,-2.4] \,  (-2.6)$} & $94\pm16$  & $12\pm2$  & $3.6\pm0.8$ & $0.23\pm0.08$ 
      \\ \hline 
      { $[-2.4,-2.0] \,  (-2.2)$} & $158\pm27$  & $27\pm5$  & $8.0\pm2.0$ & $0.68\pm0.17$ 
      \\ \hline 
      { $[-2.0,-1.5] \,  (-1.75)$} & $176\pm27$  & $28\pm5$  & $12.9\pm3.4$ & $1.34\pm0.32$ 
      \\ \hline  
      { $[-0.5,+1.0] \,\,(0.25)$} &      -        &     -   & $16.9\pm3.5$ & $1.97\pm0.35$ 
      \\ \hline 
    \end{tabular}
  \end{center}
  \parbox{16cm}{\caption{\label{tab1} Inclusive $\pizero$ photoproduction
      cross sections $d^2 \sigma_{\gamma p} / dp_\perp^2 dy$ 
      ($\mu$b/(GeV/$c)^2$) for bins in the $\pizero$ transverse 
      momentum $p_\perp$ and rapidity $y$.
      The bin centres are given in addition to the interval limits.}}
\end{table}}

{\small \begin{table}[h]
  \begin{center}
   \begin{tabular}[h]{|c||c|c|c|c|c|c|} \hline
      \,
      &\multicolumn{6}{c|}{ $p_{\perp}$ [GeV/c] } \\ \cline{2-7}
      { \raisebox{0.0ex}[-5.5ex]{$y$} }  
      &{$[0.2,0.6]$} 
      &{$[0.6,0.9]$}
      &{$[0.9,1.2]$}
      &{$[1.2,1.5]$}
      &{$[1.5,1.8]$}
      &{$[1.8,2.1]$}  
      \\  
      & (0.34)
      & (0.73)
      & (1.03)
      & (1.33)
      & (1.63)
      & (1.93)
      \\ \hline \hline
      {$[-3.5,-1.5]$} &$114\pm20$&$12.5\pm2.4$&$2.0\pm0.4$&$0.47\pm0.15$& 
      $0.14\pm0.05$&$0.04\pm0.01$\\ \hline 
    \end{tabular} 
  \end{center}
  \parbox{16cm}{\caption{\label{tab2} Inclusive $\pizero$ photoproduction
      cross sections $d^2 \sigma_{\gamma p} / dp_\perp^2 dy$ 
      ($\mu$b/(GeV/$c)^2$) for bins in the $\pizero$ transverse 
      momentum $p_\perp$ and for the rapidity range $-3.5\le y\le -1.5$.
      The bin centres are given in addition to the interval limits.}}
\end{table}}
The measurements presented in Fig.~\ref{dsint} have been performed 
in the backward
region at large negative rapidities where phase space effects
begin to be visible, causing
a damping of transverse momentum spectra. This leads to
a reduction of the cross section and to a steepening
of the $p_\perp$ distribution.  
The power law ansatz does, nevertheless,
satisfactorily describe the shape of the spectrum if the parameter
$p_{\perp 0}$ is fixed to the value found in \cite{h1cha}, namely
$p_{\perp 0}~=~0.63$~GeV/c. A fit with the power law ansatz
yields $n~=~8.0~\pm~0.2$ ($\chi^2/\mbox{ndf}~=~0.85$).
% where the error of the 
% fit reflects the errors on the data points.
%% which are dominated by systematics. 
The value is slightly larger than
the values found in the fits to the charged particle spectra in the 
central rapidity region, $n = 7.1 \pm 0.2$ 
\cite{h1cha}\footnote{In Ref.~\cite{h1cha} the error on the power $n$
is erroneously quoted as 2.0.},
 $n = 7.03 \pm 0.07$ \cite{h1cha2} (for $p_\perp > 2$ GeV/c) and 
$n = 7.25 \pm 0.03$ \cite{ZEUS} (for $p_\perp > 1.2$ GeV/c, 
with $p_{\perp 0}~=~0.54$~GeV/c).

The measured $p_\perp$ distribution in Fig.~\ref{dspt}
can also be described by an exponential distribution of the form 
\begin{equation}
\frac{d^2 \sigma}{d p_\perp^2 dy} =
   a \cdot \exp(-b \cdot \sqrt{p_\perp^2 c^2+ m_{\pizero}^2 c^4}).
\end{equation}
The fit to the data in the range $0.3 \le p_\perp \le 1.6$~GeV/c is shown 
in Fig. 4a and yields the value
$b~=~(5.7~\pm~0.3)$~GeV$^{-1}$  ($\chi^2$/\mbox{ndf}~=~1.0).
Such slope values are typical of soft hadronic interactions 
and expected in
%  thermodynamical models\footnote{From the value of 
%the slope $b$ an ``interaction 
%temperature'' of $(??? \pm ??)$ MeV can be derived.} 
thermodynamical models (see e.g. \cite{hagedorn}).
Above $p_\perp \approx 1.5$~GeV/c, however, 
         the exponential form tends to fall below the data, in contrast
         to the power law ansatz. This observation can
         be interpreted as an indication of the onset of hard 
         parton-parton scattering processes which lead to a hardening
         of the transverse momentum distribution.

This interpretation is supported by a comparison of the data
with the MC models.
Fig.~\ref{dsint}b shows again the cross section as a function of $p_\perp$,
this time together with the model predictions of PYTHIA and PHOJET. The
agreement between the data and the PYTHIA prediction 
is in general slightly better than for PHOJET. Also shown
is the PYTHIA prediction for the amount of 
``direct'' photon interactions. The contribution of  
this direct component 
accounts for $\sim 5$\% of the cross section 
at the lowest $p_\perp$ values, increasing up to  
$\sim 20$\% at the highest $p_\perp$ values accessed. 
Thus, in the rapidity and transverse momentum range covered by these
measurements, the cross section is dominated by ``resolved'' 
photon processes.

\begin{figure}[H]
  \begin{center}
    \begin{tabular}{cc}
      \hspace*{-0.0cm}
      \epsfig{file=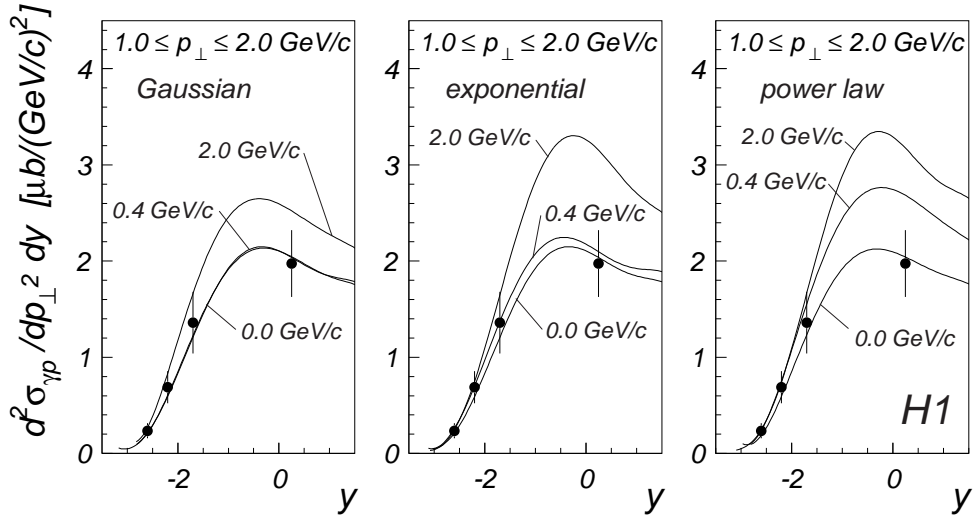,width=0.8\textwidth}
    \end{tabular}
    \parbox{16cm}{\caption{\label{chi2fits} The $ \pi^0$ cross section as 
    a function of rapidity $y$ for the transverse momentum range 
    $1.0 \le p_\perp \le 2.0$~GeV/c. 
    The curves are the predictions of PYTHIA, labelled
    with the corresponding $k_{\perp 0} $ values, as described in the 
    text. No  additional cutoff on $k_\perp$ is applied (see text).
        }} 
  \end{center}
\end{figure}

The generator PYTHIA has also been used to search for 
evidence for a non-zero intrinsic transverse momentum,
%The manifestation of a possible intrinsic transverse momentum
$k_\perp$, of partons inside the photon.
% (in its hadronic phase) 
As mentioned in section~3, PYTHIA offers the choice of
several distributions for the parameter $k_\perp$: a Gaussian, an
exponential, and a power law ansatz. All three parametrizations
have been used and the parameter $k_{\perp 0}$, which defines their  
widths, varied between $0$ and $2$~GeV/c. 
%The results have been quantified by fits to the data. 
The largest influence on the predicted shape of the $y$ distribution
appears in the highest $p_\perp$ interval. This interval, together
with the curves predicted by PYTHIA for the Gaussian, exponential, and
power law parametrizations, is shown in Fig.~\ref{chi2fits}.
We note that for these plots no additional cutoff for the primordial 
$k_\perp$ has been applied (see section~3).
%
% %%this part will be changed%%
%For the Gaussian and exponential functions large effects 
%are seen when varying the parameter $k_{\perp 0}$, and the 
%agreement with the data is worse for large values of $k_{\perp 0}$. 
%In the case of the power law function, no big variation is seen when
%varying $k_{\perp 0}$. This can be understood from the fact that when
%using the power law function PYTHIA applies for the primordial $k_\perp$
%a cutoff,
%the value of which is taken from the transverse momentum
%of the hard scattering subprocess. Thus, in the case of this
%distribution, no conclusion can be drawn on the size
%of $k_{\perp 0}$. In the case of the Gaussian and 
%exponential functions, low values of $k_{\perp 0}$ are preferred.
%
For all three functions large effects  are seen when varying the
parameter $k_{\perp 0}$, and low values of $k_{\perp 0}$ are
preferred.
Even a vanishing $k_{\perp 0}$ is consistent with the data.

Finally the behaviour of the distribution 
of the Feynman variable $x_{\rm F}$  
has been investigated, with $x_{\rm F}$ defined as 
the fraction $p_L/p_L^{max}$,
where $p_L$ is the pion's momentum
component parallel to the beam axis in the $\gamma p$ CM frame,
and $p_L^{max}$ its maximum value in a
given event, namely $p_L^{max} \approx \sqrt{s_{\gamma p}}/2$. 

\begin{figure}[H]
  \begin{center}
    \begin{tabular}{cc}
      \hspace*{-1.3cm}
      \epsfig{file=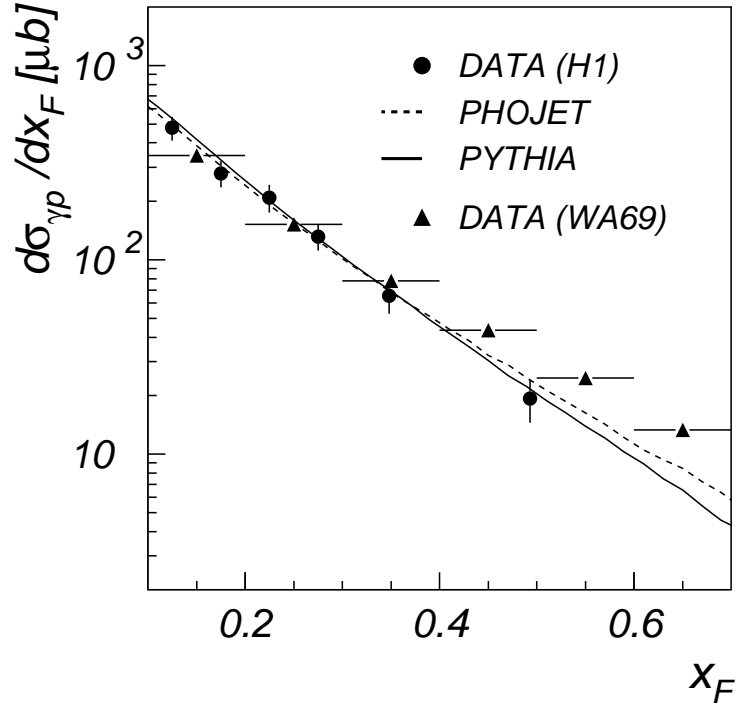,width=0.6\textwidth} 
    \end{tabular}
    \parbox{16cm}{\caption{\label{dsxf} Inclusive $\pi^0$ 
                   photoproduction cross section as a function of
                   Feynman $x$. The curves are the predictions of the
                   PHOJET (dashed) and PYTHIA (full) models. The H1 data, 
                   at $\sqrt{s_{\gamma p}} \approx 208$ GeV,
                   as well as the MC model predictions, are obtained
                   in the rapidity range $-3.5\le y \le -1.5$. The data of
                   the Omega Photon Collaboration (WA69), 
                   at $\sqrt{s_{\gamma p}} \approx 12.3$ GeV,
                   were derived as described in the text.}}
                   
  \end{center}
\end{figure}

In Fig.~\ref{dsxf}, the differential cross section
 $d\sigma_{\gamma p}/dx_{\rm F}$ is 
displayed, obtained in the rapidity range $-3.5 \le y \le -1.5$.
There is good agreement with the model predictions of PYTHIA and
PHOJET. The cross section values are given in Table~\ref{tab3}.

{\small \begin{table}[h]
  \begin{center}
    \begin{tabular}[h]{|c||c|c|c|c|c|c|} \hline
      \,
      &\multicolumn{6}{c|}{ $x_{\rm F}$} \\ \cline{2-7}
      { \raisebox{0.0ex}[-2.5ex]{$y$} }  
      &{ $[0.1,0.15]$ } 
      &{ $[0.15,0.2]$ }
      &{ $[0.2,0.25]$ }
      &{ $[0.25,0.3]$ }
      &{ $[0.3,0.4]$ }
      &{ $[0.4,0.6]$ }
      \\
      & (0.125)
      & (0.175)
      & (0.225)
      & (0.275)
      & (0.348)
      & (0.493)        
      \\ \hline \hline
      {$[-3.5,-1.5]$}&$477\pm67$&$279\pm43$&$209\pm35$&$131\pm19$&$65\pm12$
      &$19\pm5$\\ \hline 
    \end{tabular}
    \parbox{16cm}{\caption{\label{tab3} Inclusive $\pizero$ photoproduction
        cross sections $d \sigma_{\gamma p} / dx_{\rm F}$ ($\mu$b) in intervals 
        of $x_{\rm F}$, in the rapidity range $-3.5 \le y \le -1.5$. 
        The bin centres are given in addition to the interval limits.}} 
  \end{center}
\end{table}}

%
% additional paragraph on the comparison with Apsimon et al. data.
%

Data from the Omega Photon Collaboration (WA69, Apsimon et al., 
\cite{Omega}) are also shown in Fig.~\ref{dsxf}. 
They were obtained in 
fixed target $\gamma p$ collisions with an average photon energy of
$80$ GeV (corresponding to $\sqrt{s_{\gamma p}}=12.3$ GeV). 
From this CM energy and the $\pi^0$ mass one derives a CM rapidity
range of up to 4.5 units available for neutral pions.
These cross sections of inclusive $\pizero$ photoproduction, 
available in Ref.~\cite{Durham} 
as $E d^3\sigma /d p^3$ in bins of $x_{\rm F}$
and $p_\perp$, have here been converted to cross sections differential
in $x_{\rm F}$, by integrating over $p_\perp$ in each 
bin\footnote{In the lowest bin of
$p_\perp$ and $x_{\rm F}$, where a measurement with 80 GeV
beam energy is missing in \cite{Omega},
the corresponding measurement from the data set with average beam energy
$140$~GeV has been used.} of $x_{\rm F}$. 
The two experiments are in good agreement, 
although small differences are seen in 
shape and normalization. These differences can be qualitatively
understood from the
difference in rapidity range; the cuts $ -3.5 \le y \le -1.5$ 
(corresponding to $ -5.5 \lapprox\ y\ \lapprox -3.5$ in the $\gamma p$ CM)
reduce the H1 $\pizero$ cross
section at both large and small $x_{\rm F}$, a reduction 
which is only partly compensated for by the expected increase 
due to the rise of the total cross section 
between the two CM energies (factor $\sim1.3$ \cite{PDG}).

\section{Summary and Conclusions}

Inclusive $\pizero$ photoproduction in the photon hemisphere
has been analysed at HERA in the $\gamma p$ CM energy range 
$177<\sqrt{s_{\gamma p}}<242$~GeV, with average $\sim208$ GeV. 
Differential cross sections with respect to transverse momentum $p_\perp$,
rapidity $y$ and the Feynman variable $x_{\rm F}$ are presented.
The neutral pion data extend the measurements towards
the previously unexplored domain of
large rapidity in the photon fragmentation region,
%sensitive to the photon remnant in resolved $\gamma p$ interactions.
dominated by resolved  $\gamma p$ interactions.
This work extends previous H1 measurements performed with charged
particles. In the phase space region covered by both analyses there is good
agreement between the neutral and charged pion differential cross sections.

The differential cross section as a function of transverse momentum
$p_\perp$ shows an exponential fall at lower $p_\perp$ values
as seen in soft hadron-hadron collisions, 
but exhibits at values of $p_\perp$ larger than 1.5~GeV/c an enhancement   
which is expected for hard parton--parton scattering processes. 
The distribution in the entire $p_\perp$ range covered here is well
described by a power law ansatz.
The event generators PHOJET and PYTHIA are able
to describe the measured cross sections in this kinematical domain,
with a slight preference for PYTHIA. 
In the context of the PYTHIA model, the data are 
inconsistent with large values of an intrinsic transverse 
momentum in the photon.

\section*{Acknowledgements}
We are grateful to the HERA machine group whose outstanding efforts have
made and continue to make this experiment possible. We thank the engineers
and technicians for their work constructing and maintaining the H1
detector, our funding agencies for financial support, the DESY technical
staff for continuous assistance, and the DESY Directorate for the
hospitality which they extend to the non-DESY members of the collaboration.
A helpful communication with R. Engel is gratefully acknowledged.

%\newpage

\newpage

\newpage
\vspace*{-1.0cm}

\newpage
\vspace*{-1.0cm}

\newpage

\vspace*{-2.0cm}

\end{document}

%% file: h1autsn.tex
%   H1AUTS  Author list by names, no. of authors  350
%           status: 20/03/00   14.11.15
\noindent
 C.~Adloff$^{33}$,                %WUPP-ST                  Adloff             
 V.~Andreev$^{24}$,               %LPI -PD                  Andreev            
 B.~Andrieu$^{27}$,               %ECPL-PD                  Andrieu            
 V.~Arkadov$^{35}$,               %ZEUT-ST    10/96         Arkadov            
 A.~Astvatsatourov$^{35}$,        %ZEUT-ST     02/98        Astvatsatourov     
 I.~Ayyaz$^{28}$,                 %PARI-ST       08/88      Ayyaz              
 A.~Babaev$^{23}$,                %ITEP-PD                  Babaev             
 J.~B\"ahr$^{35}$,                %ZEUT-PD                  Baehr              
 P.~Baranov$^{24}$,               %LPI -PD                  Baranovp           
 E.~Barrelet$^{28}$
% ...
%     H1 Institutes as appearing on publications

\noindent
% ...
$ ^{23}$ Institute for Theoretical and Experimental Physics,
          Moscow, Russia \\
 $ ^{24}$ Lebedev Physical Institute, Moscow, Russia$^{e,h}$ \\
% ...
$ ^{27}$ LPNHE, \'{E}cole Polytechnique, IN2P3-CNRS, Palaiseau, France \\
 $ ^{28}$ LPNHE, Universit\'{e}s Paris VI and VII, IN2P3-CNRS,
          Paris, France \\
% ...
$ ^{33}$ Fachbereich Physik, Bergische Universit\"at Gesamthochschule
          Wuppertal, Wuppertal, Germany$^a$ \\
% ...
$ ^{35}$ DESY, Zeuthen, Germany$^a$ \\
% ...
$ ^a$ Supported by the Bundesministerium f\"ur Bildung, Wissenschaft,
        Forschung und Technologie, FRG,
        under contract numbers 7AC17P, 7AC47P, 7DO55P, 7HH17I, 7HH27P,
        7HD17P, 7HD27P, 7KI17I, 6MP17I and 7WT87P \\
% ...
$ ^e$ Supported by the Deutsche Forschungsgemeinschaft \\
% ...
$ ^h$ Supported by Russian Foundation for Basic Research 
       grant no. 96-02-00019 \\
% ...
%Supported by the Alexander von Humboldt Foundation \\
% $ ^{m}$ Foundation for Polish Science fellow \\